# A Fresh Look at the Architecture and Performance of Contemporary Isolation Platforms

Vincent van Rijn
v.j.vanrijn@student.tudelft.nl
TU Delft
Netherlands

Jan S. Rellermeyer
J.S.Rellermeyer@tudelft.nl
TU Delft
Netherlands

**Abstract**

With the ever-increasing pervasiveness of the cloud computing paradigm, strong isolation guarantees and low performance overhead from isolation platforms are paramount. An ideal isolation platform offers both: an impermeable isolation boundary while imposing a negligible performance overhead. In this paper, we examine various isolation platforms (containers, secure containers, hypervisors, unikernels), and conduct a wide array of experiments to measure the performance overhead and degree of isolation offered by the platforms. We find that container platforms have the best, near-native, performance while the newly emerging secure containers suffer from various overheads. The highest degree of isolation is achieved by unikernels, closely followed by traditional containers.

## 1 Introduction

The need for isolation on multi-tenant computing platforms is as old as the first attempts at multiprogramming [12, 22]. While strong, hardware-assisted forms of isolation in the form of hypervisors have initially dominated and fueled the first wave of cloud computing [3, 28], container platforms gained increased popularity based on the promise of a lower performance impact by achieving isolation within the operating system as opposed to full system virtualization, ultimately allowing for higher density [36]. More recently, this spectrum was complemented by isolation platforms which call themselves *secure containers* and aim at combining the productivity and ease of use of containers with the strong sandboxing model that hypervisors provide.

The actual performance and security of the different options is often poorly understood by researchers and practitioners alike. There is only a limited number of empirical studies and the few that exist [11, 15, 30, 34, 40] suffer from small sample sizes, mostly focus on comparing the performance of traditional hypervisors and containers, and do not include the more recently developed hybrid and specialized platforms.

With this paper, we want to address this gap and develop a broad and systematic understanding of the different architectures (Section 2), and how they influence the performance (Section 3) and security/isolation properties (Section 4) of the platforms.

Through extensive experimentation, we could confirm that container platforms have the best, near-native, performance. Hypervisors exhibit significant differences, but I/O and CPU-bound tasks typically perform on-par with native execution. Networking and memory subsystems always experience overhead. Secure containers particularly suffer from performance penalties in the I/O subsystem, but promising alternatives are being developed. Finally, unikernels perform well, but their performance is hard to characterize due to the various incompatibilities with workloads commonly used for benchmarking. Start-up time is generally the lowest for containers, whereas for the hypervisors it is highly dependent on the machine model. Furthermore, we measured the attack surface of the different platforms and surprisingly found that the secure containers execute more system calls against the host and therefore have a larger attack surface according to the Horizontal Attack Profile (HAP) [5]. Their security advantage can therefore only lie in the defense-in-depth. Beyond these overarching conclusions, we discuss 28 detailed findings which can help practitioners to make educated decisions on the best isolation platform for their given problem.

## 2 Isolation Platforms

We consider four categories of isolation platforms. Hypervisors represent the (nowadays) classic approach of using hardware virtualization support for isolating different virtual machines. Containers achieve isolation by re-using isolation barriers like processes that were already present in commodity operating systems, augmented with advanced namespaces which were introduced explicitly for supporting lightweight isolation. Secure containers combine some aspects of both worlds to achieve a stronger degree of isolation while being compatible with container platforms. Unikernels are aiming at providing a minimal shim for running individual applications in isolation. The following sections compare the different architectural design of commonly uses platforms, especially the newly emerging efforts that have gained increasing popularity in the cloud.

### 2.1 Hypervisors

Hypervisors make use of hardware virtualization instructions to create the image of a dedicated virtual machine for each guest. Consequently, guests run a full system stack, including a dedicated guest operating system. Instructions that guests are not able to execute need to be emulated through a hypercall into the hypervisor, which is a known source of overhead [23]. Beyond this general pattern, hypervisors can differ substantially in their architectures. In the recent years, new hypervisors like Firecracker and Cloud Hypervisor have emerged that are designed for a quick startup time in the cloud.

#### 2.1.1 QEMU and KVM.
In QEMU [4], every VM gets a process assigned that is dedicated to that VM and scheduled on the host OS like any other process. The host cannot see which processes are running within a VM, unlike with namespace-based isolation platforms such as containers. Memory for guests is provided through allocation by the host process, and is then mapped to the guest's address space using `mmap()`. The allocation can be backed by either RAM or file-backed memory. The guest sees this as its own physical memory.

At its core, QEMU uses an event-driven architecture (Figure 1 shows the host and guest domain) and continuously polls for new events and, if one is available, dispatches it to the appropriate event handler [21]. In QEMU, this main loop is implemented in `main_loop_wait()`, and handles the following types of events:

- Waiting for registered file descriptors to become available. These file descriptors get registered by various resources, such as the TAP network device, audio (ALSA), and drivers employing `virtio`.
- Run expired timers.
- Requests for invoking a function in another thread (such requests are called bottom-halves).

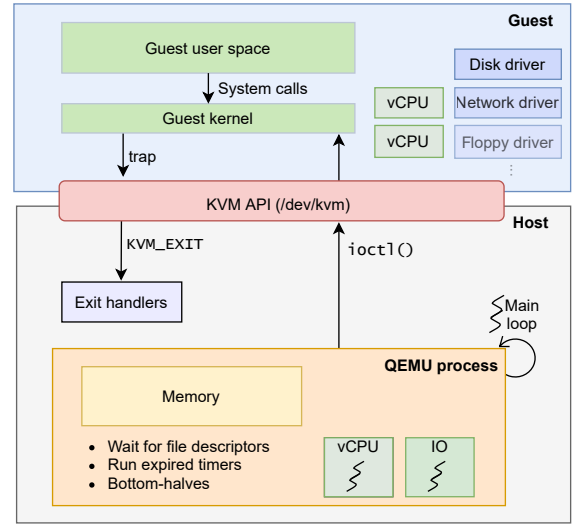

**Figure 1.** Architecture of the QEMU hypervisor, divided in a guest and host section with KVM serving as an interface between the two.

Hardware-assisted virtualization enables the native execution of guest code in a special guest CPU mode, orchestrated by the KVM [28] Linux kernel module. QEMU, when KVM support is enabled, creates and runs the KVM VM in guest CPU mode. QEMU is thus only in the loop when the guest traps out and thereby raises an event. QEMU interfaces with KVM through the special `/dev/kvm` file, and the VM and vCPUs are created through specific `ioctl()` system calls. Executing `ioctl(KVM_RUN)` (resulting in a VM_ENTRY) hands over control to the guest, and keeps it running unless the guest traps back to QEMU.

#### 2.1.2 Firecracker.
Firecracker [1] adopts an event-driven architecture and uses KVM to create and run VMs, much like QEMU. What sets Firecracker apart is its minimalist design, especially compared to QEMU/KVM, that focuses on lower overhead. In total, Firecracker supports only a handful of emulated devices, namely a set of paravirtualized `virtio` drivers (e.g. `virtio-net` and `virtio-blk`), a legacy i8042 serial and PS/2 mice and keyboard controller, and a pseudo clock device that records time since booting.

Another technique Firecracker implements to reduce boot time is making use of the Linux 64-bit boot protocol. This allows for booting directly into 64-bit mode, skipping the usual x86 mode-by-mode (from the 16-bit real mode to 64-bit long mode) booting protocol. Furthermore, Firecracker boots directly into an uncompressed Linux kernel, starting at the 64-bit entry point. This is different from typical Linux platforms, in which the kernel decompresses itself at startup.

#### 2.1.3 Cloud Hypervisor.
The main difference to the systems discussed so far is that Cloud Hypervisor [9] finds a balance between a minimal design (Firecracker) and a feature-complete full system emulation (QEMU), slightly leaning towards the minimalism of Firecracker. As such, the architectural properties of Cloud Hypervisor are expressed here in terms of how it deviates from the Firecracker design.

Cloud Hypervisor supports 16 different devices, in contrast to the 7 of Firecracker and 40+ of QEMU. The majority of the devices in this device model are paravirtualized `virtio` devices. In contrast to Firecracker, Cloud Hypervisor supports vhost-user devices, hotplugging memory, and vCPUs. Requests for hotplugging are performed via an API that Cloud Hypervisor exposes. Memory is hotplugged by first allocating it on the host (must be a multiple of 128 MiB) and then mapping it from the hypervisor userspace process to the virtualized memory of the guest. Hotplugging extra CPUs is implemented by the host performing a `CREATE_VCPU ioctl()` call, and then advertising the vCPU cores to the running guest kernel using `ACPI`. The newly provisioned vCPUs are not automatically used within the guest but have to be brought online by manual interaction with the guest Linux kernel `sysfs` interface.

While these features set Cloud Hypervisor apart from Firecracker, it is still similar in its goal of providing a lean execution environment that is optimized for a reduced startup latency.

### 2.2 Container Platforms

Containers provide a dedicated system image while not employing a separate kernel. Instead, they rely on traditional process isolation paired with namespaces as an effective and low-overhead isolation mechanism.

#### 2.2.1 Docker.
Docker refers to an entire software suite including a container lifecycle manager, packaging software, and an interface for communicating with a repository of online container images (Docker Hub). Broadly speaking, Docker uses a client-server architecture. The CLI client is what an end-user interacts with through familiar commands (e.g. `docker run`). These commands are sent to the Docker daemon `dockerd` which is responsible for building, running and distributing the Docker containers [13].

One crucial component in the Docker system is its runtime which handles the creation of the isolated containers. The default Docker container runtime is `runc`. This runtime, given a layered file system and related container metadata, creates a new isolated container. `runc` uses functionality exposed by the Linux host kernel to enforce isolation between a container and the host operating system. The kernel is thus shared between the host operating system and the container, and no new kernel is booted. The two main kernel features that are core to `runc`'s isolation are namespaces for reducing visibility of kernel resources from a container and cgroups to constrain the available system resources to a container.

#### 2.2.2 LXC.
Linux containers (LXC) approach the implementation of containers in a way similar to `runc`, using namespaces and cgroups as the main isolation mechanisms.

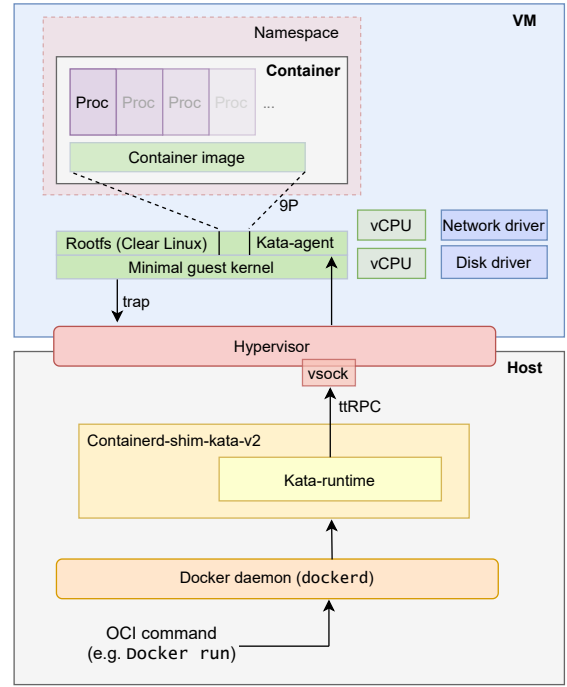

**Figure 2.** Architecture of the Kata containers secure container platform. The guest section consists of a namespaced container running within a hypervisor-based VM, in which the workload is run.

In fact, up until a year after the release, Docker used LXC as library (`liblxc`) to set up its containers (but now uses its own separate re-implementation `libcontainer`). The characteristic that sets LXC apart from Docker is its ability to create an environment as close as possible to a standard Linux installation, while still avoiding the need for a separate kernel. Concretely, this means that LXC containers implement a fully-fledged `init` system like systemd, whereas Docker uses a simplified system. LXC also makes use of the feature-complete ZFS file system, instead of a layered file system like Docker (Docker does have a ZFS storage driver, but its use is discouraged [1]). It is worth mentioning that LXC already provides the user with a way to run non-root unprivileged containers, making use of the newer cgroups v2. Docker, at the time of writing, only offers running containers using root privileges.

### 2.3 Secure Containers

While traditional containers solely rely on software-based isolation, secure containers are hybrids in the sense that they use hardware-based isolation for critical components.

#### 2.3.1 Kata containers.
Kata containers [35] combine a hypervisor as the core isolation mechanism with namespaces to achieve the usability of a container platform. As Figure 2 illustrates, the entry point of the Kata containers architecture

---
[1] https://docs.docker.com/storage/storagedriver/zfs-driver/

is `kata-runtime`, which the user interacts with through the Docker daemon.

The `kata-runtime` component is responsible for starting the hypervisor, which requires a kernel and a root file system to start. Shipping with the `kata-runtime`, there is a Linux kernel that "is highly optimized for kernel boot time and minimal memory footprint" [27]. This optimization in practice boils down to disabling nearly all kernel features for the guest kernel using `kconfig`. The runtime passes a 'mini OS' as its root file system. This mini OS is customizable while building from source, but by default it is based on Clear Linux [8]. It uses `systemd` to start the `kata-agent` immediately.

The `kata-agent` is a process for managing containers and processes running within a hypervisor. This agent communicates with `kata-runtime` using a ttRPC server (a re-implementation of gRPC specifically for low-memory environments [39]) that is exposed on the host by the hypervisor through a vsock file. A confined (namespaced and cgrouped) context is created by the `kata-agent` within the hypervisor. The root file system of this newly created confined context is that of the original Docker image, passed as a shared mount point from the host through the hypervisor. Other settings, such as which command should be run at start, are passed to the `kata-runtime` through the Docker image. This is set up within the new container context, and the workload is run. Whenever a `docker exec` statement is issued to `kata-runtime`, and a Kata container is set up already, it simply forwards this command to the `kata-agent` running inside the hypervisor, which delegates it to the confined context to create a new process with this new command.

**2.3.2 Google's gVisor.** gVisor [20] takes a different approach in which no hypervisor is used. Instead, system calls in gVisor are intercepted and redirected through use of a 'platform'. Concretely this platform leverages either `ptrace` or KVM. The `ptrace` system call interception implementation employs `PTRACE_SYSEMU` to stop and intercept the execution of system calls into the host kernel. With KVM as the platform, the main gVisor process is run as a KVM VM. In general, the KVM mode ought to be faster because `ptrace` has a relatively high context-switch penalty while KVM can make use of hardware assisted virtualization features like fast address space switching [19].

Figure 3 illustrates the general architecture of gVisor. Regardless of which platform is used, system calls get intercepted and consequently bounced back to a particular process in user-space. This process is called the `Sentry`. The `Sentry` is a kernel in user-space, implementing not just system calls but also features like signal delivery, memory management and the threading model. To reduce the attack surface, the system calls in the `Sentry` process are implemented using a small subset of system calls to the host kernel. This is enforced through seccomp filters. The `Sentry` process itself

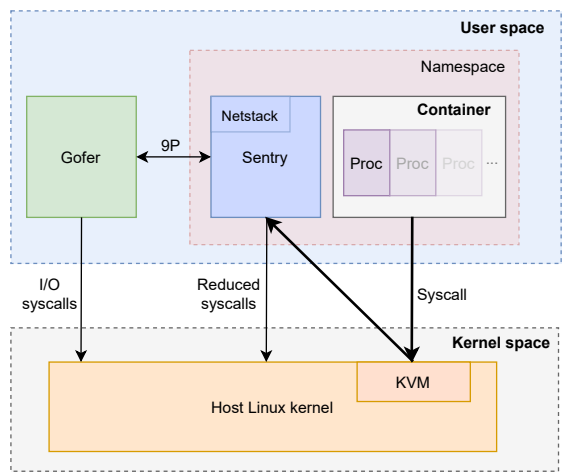

**Figure 3.** Architecture of the gVisor secure container platform. System calls from a gVisor container get redirected to the `Sentry` component (using KVM). I/O system calls are handled by the `Gofer` component.

runs as an unprivileged process and limits its view of the system through namespaces.

The underlying idea in gVisor is that there is defense in depth. Not only does the `Sentry` process re-implement system calls to reduce the attack surface, it also runs within its own namespace. Even if the `Sentry` process were to be compromised, the attacker would have to break out of the namespaces. The seccomp filters applied to the `Sentry` also include all I/O related system calls. This means that the `Sentry` cannot dispatch any I/O related system calls to the host kernel but instead needs to dispatch them to another gVisor component called `Gofer`. The `Sentry` and `Gofer` process communicate via the 9P protocol originally developed for the Plan9 operating system [33], similar to how the file system is shared between the hypervisor guest and host in Kata containers.

### 2.4 Unikernels

The fundamental idea of single-address space operating systems with a minimal set of APIs goes back to the original work on the Exokernel [14]. Recently, the idea has gained traction again in the form of library operating systems for virtualization where the virtual machines operate on top of libraries which interface with the host system as opposed to a full independent OS kernel [6, 31]. While a comprehensive comparison is beyond the scope of this paper, we include OSv in our study since it is a supported isolation option by Firecracker.

**2.4.1 OSv.** OSv [29] is a unikernel that uses existing compilers and a custom kernel to call into. A schematic overview of the architecture of OSv is given in Figure 4. The OSv kernel includes a dynamic ELF linker that can run standard code compiled for Linux. This linker maps the executable and its dependencies to memory. Whenever application code calls functions from the Linux ABI through the standard C library, the linker dynamically resolves it to the corresponding function implemented by the custom OSv kernel. This means that system calls, called through the wrappers implemented in `glibc`, are treated as regular function calls, and do not lead to a user-to-kernel mode switch. Both the application and kernel (i.e. OS library) run in the privileged ring 0.

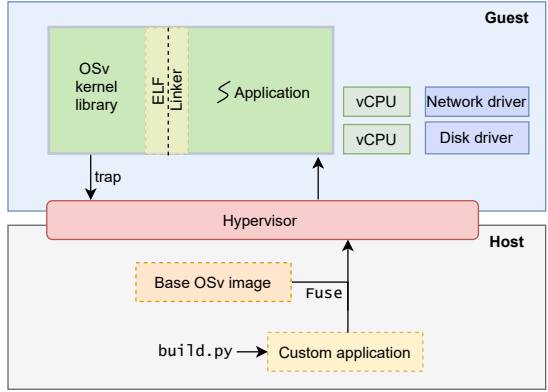

**Figure 4.** Architecture of the OSv unikernel. The unikernel image is built by fusing the application and a base OS, and is run using a hypervisor. The custom ELF linker enables the circumvention of mode switches.

Running a unikernel image on OSv is done through existing hypervisors. OSv images consist of a base image that is fused together with cross-compiled source code that calls into this base image. The OSv base image exposes an interface that follows the Linux ABI convention. This setup makes it so that as long as executables are compiled as a relocatable shared object (`.so` in Linux) as well as in the form of a position-independent ('PIE') binary, recompilation and thus application source code is not required, and existing code can transparently call into the OSv kernel. One limitation remains, as is typical for unikernels: there is no support for multiple processes within one guest, and as such, system calls like `fork()` and `exec()` are not available. This effectively limits potential applications to multi-threaded architectures and prohibits the use of multi-process applications.

## 3 Performance Study

It is commonly assumed that containers provide a more lightweight alternative for isolation than traditional hypervisors. Beyond this, however, little is known how much impact the different ways of virtualization have on critical components like compute, memory, network, storage, or the initial startup latency. We therefore conducted a broad set of experiments on different popular systems to develop a more systematic understanding of their performance properties. Furthermore, we evaluated the performance of several real-world workloads to get a more comprehensive picture. All experiments were conducted on a dual-socket AMD EPYC2 7542 CPU setup with 64 threads each, 256 GiB of RAM, a dedicated fast NVMe SSD as storage, and running Ubuntu Linux Server 20.04 LTS. The full benchmark setup is available on GitHub[2]. All runs were, unless noted otherwise, repeated at least 10 times, the graphs show the average over the runs and the error bars show the standard deviation.

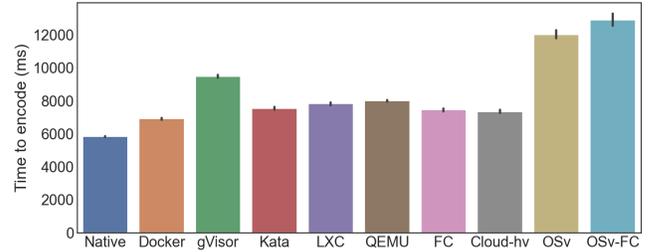

**Figure 5.** ffmpeg video re-encoding CPU bound benchmark, re-encoding a 1080p 30Mb video from H.264 to H.265. Time in ms, per platform.

### 3.1 Compute

For CPU performance, we are focusing on workloads that are primarily compute-bound. The particular benchmarks we have chosen are a video encoding task and a simple prime verification algorithm benchmark. The video encoding task entails loading a 30MB video file[3] into memory, and then encoding that video file from H.264 to the H.265 video codec. For this benchmark we make use of the `ffmpeg` [16] program. The task is executed on guests that have access to 16 CPU cores, and the job itself is executed using 16 threads.

By making use of the different presets that `ffmpeg` exposes, we have an instrument to control the speed at which the media is encoded at. In this experiment, we have used the 'slower' preset which trades CPU cycles for a higher compression ratio. The difference in performance between platforms in this benchmark can thus be attested to actual differences in CPU performance, not I/O (which we found to be the cause of significant variations between platforms in initial runs of this benchmark).

As Figure 5 shows, most of the runs end up at around 65000 milliseconds, while some differences between platforms can be observed. A surprising outlier is OSv, taking up significantly more time to re-encode the video file. Since `ffmpeg` uses a multi-threaded architecture, we suspected the difference in thread scheduling between OSv and the other platforms to be the main source of overhead. Moreover, we suspect that execution of complex SIMD instructions in the more experimental platforms induce overhead as well. In order to corroborate this hypothesis, we carried out second (single-threaded) microbenchmark. This benchmark is part of the Sysbench [37] CPU benchmark, and verifies whether a number is prime. In this experiment, every isolation platform, including OSv, performed nearly equivalently, asserting that the CPU overhead is not inherent to any of the isolation platforms.

---

[2]https://github.com/rellermeyer/container_benchmarks.git
[3]Downloadable from the Blender project website at https://peach.blender.org/download/

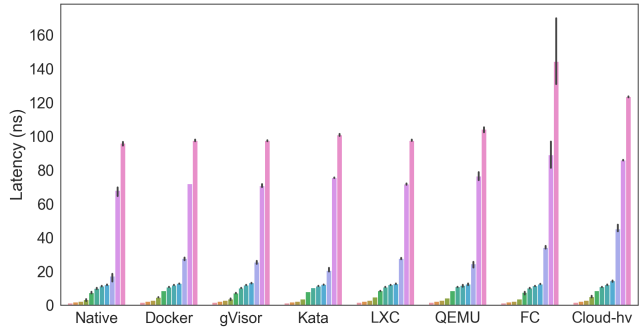

**Figure 6.** Memory latency tinymem benchmark showing the average time for accessing a random element within buffers of increasing sizes ($2^n$ from $n = 16$ through $n = 26$). The larger the buffer, the higher the latency, due to an increasing proportion of TLB cache misses.

**Finding 1:** We confirm that for CPU-bound workloads that exercises a basic subset of all available CPU instructions there is no performance overhead. However, with more complex CPU-bound tasks, such as re-encoding a video using ffmpeg, differences in performance overhead become apparent. In particular the platforms that implement custom thread schedulers (e.g., OSv) appear to experience a severe performance penalty.

### 3.2 Memory

For the memory subsystem, we evaluate both throughput and access latency. To that end, we employ the Tinymembench [38] and STREAM [32] memory benchmarks.

Tinymembench is a relatively simple benchmark which reports both the maximum bandwidth achieved through sequential memory accesses as well as the latency of random memory accesses in increasingly larger buffers. Figure 6 shows the average time for accessing a random element within buffers of increasing sizes. We can see that the larger the buffer is, the higher the latency. This is due to the increasing proportion of accesses that miss the TLB cache and need to be dispatched to L1/L2 cache, and for even larger buffers to SDRAM. The numbers displayed here indicate the extra time that was needed on top of the L1 cache access latency. The latencies of access to HugePages are omitted because both Kata containers do not support them, and more importantly, the relative results of the various platform compared to one another are almost equal to those of regular sized pages shown above. In absolute values, we recorded significant speedups in access latencies to HugePages. In general, the larger the buffer, the more it benefits from HugePages. The average access latency shrunk equally across all platforms that support HugePages by nearly 30% in the larger buffers.

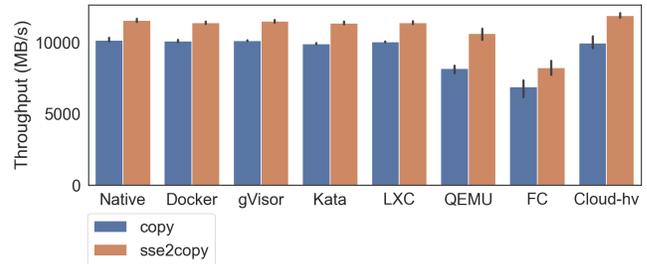

**Figure 7.** Memory throughput tinymem benchmark. Results show the number of sequential bytes copied per second using both regular as well as SSE2 instructions.

The results sketch an outcome that is mostly consistent between all the platforms, with the exception of the hypervisors. In particular Firecracker both has a higher average latency as well as standard deviation for accesses in larger buffers. The average access latency for Cloud Hypervisor is larger as well, but not to the same extent as Firecracker. These two platforms, Firecracker and Cloud Hypervisor, make use of the vm-memory Rust crate dependency, which provides a hypervisor-agnostic interface to physical memory for the virtualized guest. This dependency is responsible for fundamental memory operations such as allocation and (virtual to physical) guest memory address translation, and is therefore likely to be the cause of the higher access latencies observed for these two platforms..

We use two benchmarks to measure memory throughput. The first benchmark, in Figure 7, shows how many bytes can be copied per second using both regular as well as SSE2 instructions. This benchmark is part of the Tinymembench benchmark. The second experiment uses the popular STREAM benchmark, a simple synthetic benchmark for measuring sustained memory bandwidth by performing simple operations on vectors [32]. The STREAM benchmark consists of 4 different vector operations, but we only present the COPY operation results here as the operations yielded similar relative performance. The COPY benchmark executes code of the form a[i]:=b[i], transferring 16 bytes per iteration, and executes no floating point operations. Both of these benchmarks have a sequential access pattern, meaning that performance is minimized by memory bandwidth rather than latency (as hardware typically prefetches the data that will be requested later on).

In Figure 7 and Figure 8 we see the results of these benchmarks. The throughput performance is reminiscent of the latency plot. All platforms generally perform close to equal, with the hypervisors underperforming. Although paravirtualization and hardware-assisted virtualization have substantially reduced the overhead of hypervisors, the additional layer of hypervisor indirection seems to cause overhead. However, there is one remarkable result that contradicts this finding. Kata containers uses a hypervisor, yet is is not victim to the reduced memory latency and throughput. We thus conclude that the overhead is not inherent to the use of hypervisors. Concretely, Kata containers avoids this virtualization penalty by techniques such as the QEMU NVDIMM feature, which provides a memory-mapped virtual device that directly maps between the VM and host, bypassing the intermediary virtualized layer. Another technique that can provide improved memory performance for virtualized guests is Kernel Samepage Merging (KSM) [2]. KSM enables the sharing of memory between multiple processes (like VMs), which increases density, and therefore the reuse of hot pages (for a higher cache hit ratio). Although direct access techniques such as the NVDIMM feature and KSM potentially lead to performance gains, it also weakens the isolation boundary between tenants of the same host (as shown in e.g. [25], in which the authors present a vulnerability introduced by KSM).

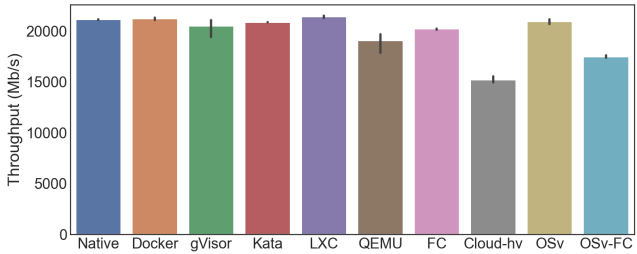

**Figure 8.** Memory copy throughput STREAM benchmark, sequentially transferring 16 bytes per iteration, total allocation size of 2.2GiB. Throughput shown is the average of the maximum of 10 runs.

In summary we can observe that most of the isolation platforms do not impose significant overhead on the use of the memory subsystem, of which we have quantified both access latency as well as throughput.

**Finding 2:** All containers, including secure containers, perform on-par with native for CPU-bound tasks.

**Finding 3:** Although most hypervisor-based platforms exhibit some form of slowdown in both latency and throughput, the Kata container platform is not significantly impaired, despite its use of the QEMU hypervisor. Furthermore, the OSv platform running under QEMU also does not show any slowdown. As such, we can conclude that the usage of a hypervisor does not unconditionally lead to memory performance overhead.

**Finding 4:** The memory performance outlier is Firecracker, scoring substantially lower than the other platforms. Cloud Hypervisor shows a similar (although weaker) effect on memory access latency but not for throughput, while the opposite holds for QEMU. This suggests a trade-off between latency and throughput for general hypervisor-based platforms.

**Finding 5:** OSv's memory performance is strongly affected by its hypervisor. OSv running under the Firecracker hypervisor underperforms in comparison to OSv running under QEMU, which yields results close to native.

### 3.3 I/O

For benchmarking the I/O subsystem we use the Flexible I/O tester (`fio`) benchmarking tool [17], version 3.25. Specifically, we use `fio` to benchmark the block I/O performance of the different isolation platforms. By benchmarking on the block level, rather than on the file system level, we solely capture the overhead imposed by the actual virtualization mechanisms, rather than a combination of overhead imposed by both the file system and block layer.

Fio measures the average read and write throughput by pre-allocating a file two times the size of the amount of memory available to the platform, using `fallocate()`, and then uses this file to write to and read from in blocks of 128kb using the `libaio` I/O engine. For these benchmarks, we start the platform that is being tested, and then attach a separate storage medium through the user interface exposed by that platform. For Docker this could be as simple as passing a bind mount through the `--volume` flag, whereas for LXC this entails creating a new ZFS storage pool on the separate storage medium and recreating the LXC container within this new pool. For hypervisors, the target storage medium is attached as an additional drive and mounted within the guest. The amount of data read/written for each I/O test is equal across the different platforms, in order to keep the chance of anomalous seek times imposed by a larger test file to a minimum. Since Firecracker does not support attaching extra storage devices, it is excluded for this benchmark. For OSv there is no working implementation of the `libaio` engine, and picking other I/O engines leads to either an unfair comparison or an underutilization of the actual I/O throughput.

In Figure 9 we see the results of this 128kb write/read throughput benchmark. Generally speaking, the read performance of Docker, LXC and QEMU/KVM is equal to that of a native platform without virtualization. The write speeds of these platforms come close to that of native as well, although overhead comes in the form of a higher standard deviation. The other hypervisor, Cloud Hypervisor, performs significantly worse, with lower throughput in both read and write performance as well as standard deviation. This however is not inherent to the use of hypervisors, as QEMU demonstrates. The secure containers gVisor and Kata containers suffer severely from the extra layers of indirection, in the best case reaching only half of the speeds achieved using other isolation platforms.

We found the I/O benchmarks to be the most difficult to run, due to caching problems with the chosen isolation platform. It is particularly difficult with hypervisors where both the guest and host have a separate buffer cache. I/O Benchmarking tools like Fio support writing directly to storage, effectively bypassing the buffer cache (using the `direct=1` option). With some isolation platforms, however, there are two separate kernel instances running (the host and guest kernel), both utilizing their own buffer cache. Despite Fio being instructed to write directly to storage and skip the buffer cache, this only circumvents the guest kernel cache. The root file system of the guest is presented as a block device to the guest by creating a loop device on the host, and flags like direct are not propagated properly. As a result, I/O requests executed within the guest can still be cached inside the host buffer cache. This can lead to incorrect benchmark outcomes, in which hypervisors outperform the native I/O speeds by a large margin. An effective way to remedy this issue is explicitly dropping the buffer cache on the host manually before each benchmark run.

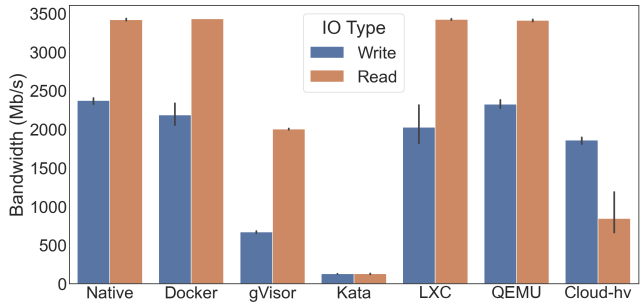

**Figure 9.** Fio I/O throughput benchmark for all platforms, excluding OSv and Firecracker. Writes and reads are in blocks of 128kb using the libaio I/O engine.

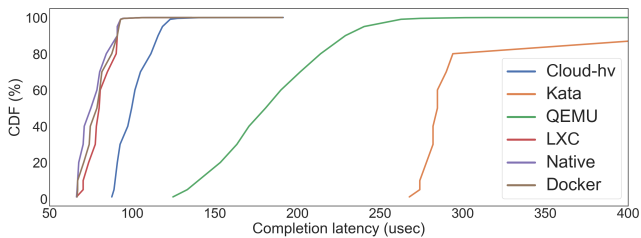

**Figure 10.** Fio randread latency benchmark for 4kb sized blocks (libaio).

Most of the platforms roughly achieve the same throughput, leaving gVisor, Kata containers and Cloud Hypervisor behind. When considering the random read latency for the platforms, of which the results are indicated in Figure 10, the relative performance of the platforms is mostly consistent. Reads are in blocks of 4kb. The hypervisors incur a latency issue that is inherent to the extra virtualization layer. QEMU experiences overhead similar to what is shown in prior research. The newer Cloud Hypervisor platform performs remarkably well in this latency benchmark, but considering its poor throughput performance, it cannot be concluded that it overall performs better than the QEMU I/O subsystem. Among the secure containers, Kata containers perform exceptionally poorly. Although the gVisor platform is excluded in this particular benchmark as all its reads got cached even when both host and guest page caches were dropped, it is reasonable to assume that its performance would be similarly lackluster due to the reliance on the 9P file system.

Although this file system is a mature piece of software by most standards, active development ceased in 2012. With the increasing interest in containers in industry, the need for a better and more performant shared file system became clear. This led to the creation of virtio-fs [4], a file system implemented in FUSE using virtio as the transport layer. Since hosts and guests of isolation platforms are not physically separated by a network, an assumption that traditional networked file systems are built upon (such as the 9P file system), no longer hold. The virtio-fs file system can take advantage of these new conditions, and can gain a significant performance speedup relative to existing networking file systems. We have carried out additional experiments, and found that Kata containers with virtio-fs significantly outperforms 9P, and is on-par with the performance of the QEMU platform in this section.

In conclusion, we have seen that quantifying I/O performance for the various isolation platform proved to be difficult due to the various layers at which caching happens.
**Finding 6:** The I/O performance of most systems is close to native except for the secure containers gVisor and Kata containers, and for the hypervisor Cloud Hypervisor.
**Finding 7:** For Kata containers, virtio-fs is a promising alternative that significantly outperforms the older 9P protocol.
**Finding 8:** gVisor performance, in its current form, is severely hampered by the use of both the 9P protocol and separate Gofer architectural component.
**Finding 9:** Cloud Hypervisor should get better as it matures, but for now remains the outlier. For this platform, there should not be an architectural bottleneck, as QEMU performs close to native.

### 3.4 Network

For measuring the network bandwidth we have used the iperf3 benchmark [24], in which the host acts as client to the server that is run inside the virtualized guest. For measuring the native benchmark performance, the host machine acts as a server while the client requests are sent from a device that is directly connected to its NIC. The iperf3 benchmark aims to reach the maximum achievable throughput over an IP network. In this context that implies that any score below native indicates overhead from the platform used.

From the results (Figure 11) we can conclude, unlike in the memory benchmark, there is always a price to be paid for virtualization (or isolation). The host achieves a mean throughput of 37.28 Gbit/s whereas the second highest, OSv, achieves a mean throughput of 36.36 Gbit/s. The performance advantage of OSv, which runs under QEMU, and a plain QEMU guest is a very significant 25.7% (in the figure, QEMU vs. OSv). The network performance throughput as exhibited here does not necessarily reflect a superior architecture of OSv, however, as the results of the benchmark using the Firecracker hypervisor to run an OSv guest only results in a less significant 6.53% increase (in the figure, Firecracker vs. OSv-FC).

---
[4]https://virtio-fs.gitlab.io/

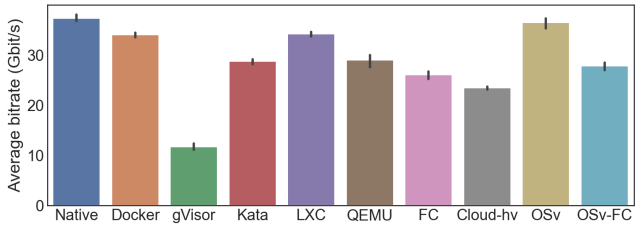

**Figure 11.** iperf3 network throughput benchmark, maximum throughput achieved over 5 runs.

The mechanism used to isolate the host from the guest network, either namespacing or virtualization, seems to have a significant impact on the network performance. Docker and LXC use a network bridge approach and incur a 9.84% and 9.19% performance penalty, respectively. The hypervisors use a TAP device and `virtio-net` setup, and incur a more severe performance penalty in the order of 25%. The less mature platforms, particularly Cloud-Hypervisor suffer from severe inefficiencies in its implementation, considering the high-level architectural setup of QEMU and Cloud Hypervisor are equal. Kata container employs both bridges and a QEMU (TAP device + `virtio-net`) setup. This means that the performance of Kata containers should be equal to the performance of this weakest link, which in this case is the QEMU part of the architecture, and indeed it is. Finally, gVisor is an extreme outlier. gVisor implements its own network stack. Implementing a network stack from the ground up, however, is not a trivial task, and as a consequence, gVisor does not yet implement all RFCs related to networking, of which many also promise increasing network throughput. Although at this time it is expected that eventually all relevant RFCs will be implemented in Netstack, for now, its performance is not competitive.

**Finding 10:** The average latency as measured with the Netperf benchmark yields similar results, with the containers using bridges (Docker, Kata containers and LXC) performing very well, followed by the hypervisors.
**Finding 11:** OSv does not outperform every other platform but has slightly lower latencies than the hypervisors.
**Finding 12:** gVisor is once again a notable negative outlier, with a 90th percentile response time 3 to 4 times that of its competitors.

### 3.5 Startup Time

The startup time is an important concern for environments in which regions of isolation need to be spawned and despawned quickly, e.g., in serverless computing. In our experiment, we measure the total end-to-end process time, from process creation to termination. The process terminates itself through a patched `init()` system (for e.g. the hypervisors and LXC) or an 'exit' entry point in the containers. OSv startup time is measured by invoking it without a program to run, resulting in an immediate shutdown after it completes its boot sequence. The results in this section are based on measuring the time for booting an isolation platform 300 consecutive times. Although one might assume that including process termination in the total measurement time may not be ideal, the alternative would be to use measurement methods which can not be applied to every single platform. Moreover, in practice, the overhead for process termination was minimal (1–2%), as found out through experimentation.

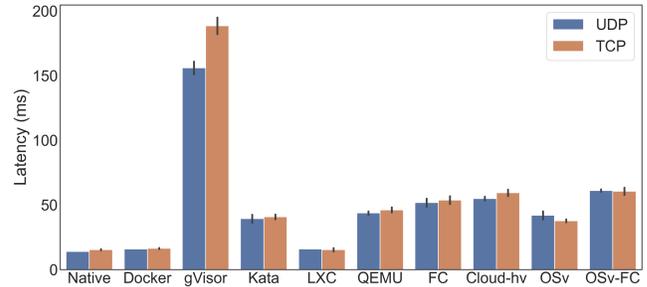

**Figure 12.** Netperf network latency benchmark (90th percentile) over 5 runs.

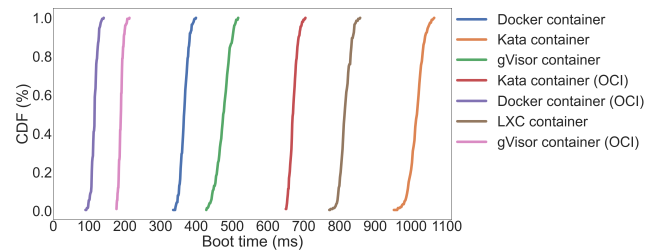

**Figure 13.** Time taken to boot container runtimes (CDF), 300 startups per platform. OCI times are obtained by directly invoking the underlying container runtime, circumventing the Docker daemon overhead.

Figure 13 shows the startup times for the examined (secure) containers. Docker is very fast to boot, taking around 100 ms, followed by gVisor at around 190 ms. Much later, at 600 ms and 800 ms, we see the Kata container and LXC platforms, respectively. The comparably slow startup time for Kata containers can be explained by the relatively complex booting sequence it has to perform, as it has to both set up namespaces as well as set up (and boot) a hypervisor. LXC, however, is an unexpected outlier. Its high booting time can be attributed to the systemd `init` system it by default uses, which takes up significantly more time than Docker's `tini`. Finally, by comparing the OCI and non OCI versions of our containers, we see that creation of containers through the Docker daemon causes a slowdown of around 250 milliseconds in startup time.

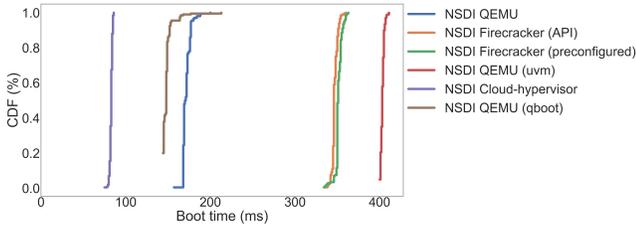

**Figure 14.** Time taken to boot hypervisors (CDF), replication of work in [1], 300 startups per platform.

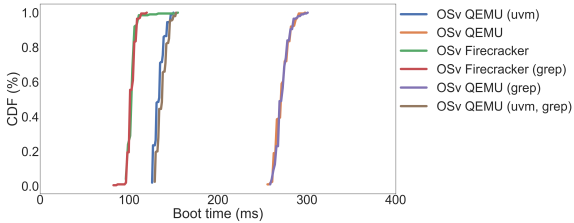

**Figure 15.** Time taken to boot OSv under supported hypervisors (CDF), 300 startups per platform.

Figure 14 shows the startup time for the different hypervisors booting with the same kernel and root file system. As mentioned before, the `init` system is patched to immediately quit as soon as it starts. The fastest hypervisor, shown at left side of the Figure, is Cloud Hypervisor, significantly outperforming the other hypervisors, followed by QEMU (both plain QEMU and the QEMU minimal `qboot.bin` BIOS firmware). The slowest hypervisor is QEMU with the µVM device model, as inspired by Firecracker. In theory this should lead to faster startup times (with fewer devices to manage, and no BIOS at startup), but in practice for this version of QEMU it only leads to increased startup time.

At around 350 milliseconds the Firecracker hypervisor appears. This is an interesting result, since in [1] the authors conclude that the FC-pre platform has the fastest startup time of all the hypervisors. We consider this conclusion skewed, as not the actual end-to-end (process creation to termination) time is measured, as it is for other platforms, but instead the time taken to write to a special device during boot time by using a patched kernel is measured. We do not consider this a fair comparison, as it does not take into account what happens before the kernel is launched, as well as anything that happens after the kernel prints this timestamp. The true time needed to start Firecracker, in our experiments, is significantly higher than its hypervisor competitors Cloud Hypervisor and QEMU.

The results in Figure 15 indicate the OSv boot times using different hypervisors. We used two main methods two measure this startup time: end-to-end (as in our measurements above) and stopping the measurement when a specific line of text is printed to `stdout`. As we can see, those two variants of each platform setup are almost superimposed on top of one another, further strengthening the credibility of our way of measuring end-to-end. An interesting observation from these results is that the results are almost opposite of the prior hypervisor results (in Figure 14): Firecracker is the fastest, QEMU µVM ranks second, and at last place we see regular QEMU.

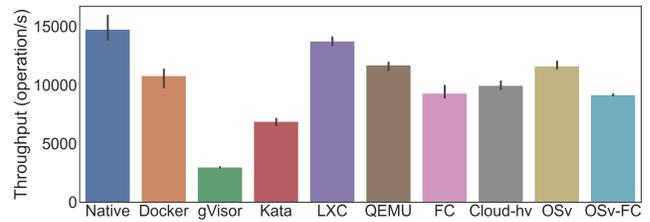

**Figure 16.** Memcached YCSB benchmark over 5 runs.

In conclusion, we find that most platforms are able to boot and exit within 200 milliseconds. We make the following observations:

**Finding 13:** Containers are fast to boot, with the exception of Kata containers and LXC, which typically take over 600 ms.

**Finding 14:** Firecracker, despite its focus on serving the serverless computing paradigm, boots the slowest out of the three hypervisors. Cloud Hypervisor is the fastest. Boot time depends heavily on the used machine model, as QEMU with the µVM machine model is (unexpectedly) the slowest out of all.

**Finding 15:** Unikernels (OSv) are faster to boot than regular Linux-based images, generally as fast as containers. Booting OSv images using different hypervisors has a significant effect on boot-time.

**Finding 16:** Measuring boot-times end-to-end (from process creation to termination) using `time` is as accurate as other ways (e.g. by checking stdout using `grep`). Overhead for this measurement technique is negligibly (1-2%) small across the various platforms.

### 3.6 Memcached

Memcached is a high-performance key-value store [18], often used as an additional caching layer in between e.g. a web server and a database. It stores small chunks of arbitrary data in memory and never materializes any of its content to disk. We benchmark the performance of Memcached on each platform using the YCSB (Yahoo! Cloud serving benchmark) [10], a popular framework for benchmarking different key-value and cloud data stores. Specifically, we use the 'workload a' preset of YCSB, a mix of 50/50 reads and writes, behavior exhibited by e.g. a session store recording recent actions.

Benchmarking using YCSB and Memcached stresses the memory and networking subsystems. As a reminder, the hypervisors underperformed in the memory and network microbenchmarks (the less mature the hypervisor the worse), and gVisor in particular did not fare well in the network microbenchmark. We see these prior results reflected in the

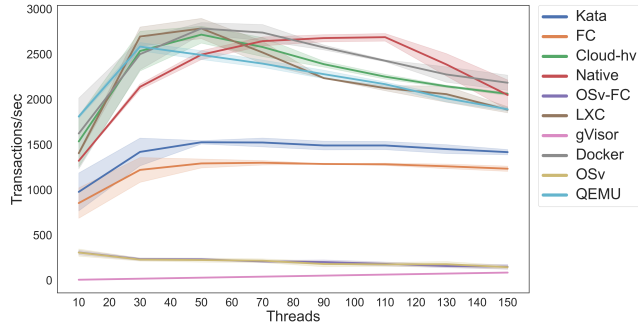

**Figure 17.** MySQL Sysbench oltp_read_write benchmark with increasing threads, over 3 runs.

Memcached benchmark in the benchmark results in Figure 16.

**Finding 17:** The newer hypervisors perform worse, and overall the regular containers (in particular LXC) perform very well.

**Finding 18:** The result of Kata containers is surprising, since the microbenchmarks of network and memory for Kata containers would not suggest a score significantly lower than most of the other platforms.

**Finding 19:** The gVisor Memcached score, although poor, can be attributed to its network performance.

### 3.7 MySQL

MySQL is a well-known relational database. In combination with the Sysbench [37] benchmarking tool (version 1.0.20), we stress the isolation platforms running MySQL version 5.6.45 using the `oltp_read_write` benchmark. This benchmark stresses the memory, file system and networking subsystems. It initially stores 1 million records into 3 tables, and then consecutively executes a SELECT, UPDATE, DELETE and INSERT SQL query. We call the combination of one of each of these queries a transaction. We perform this benchmark for every platform with an increasing number of threads, starting at 10 up until 160 threads.

The results of this benchmark are shown in Figure 17. There are numerous interesting observations to be made from these results:

**Finding 20:** For nearly all platforms, the number of transactions per second peaks at around 50 threads, after which thread contention appears to impair overall performance. The native platform peaks at around 110 threads instead, yet does not deliver a significant performance increase over the isolation platforms.

**Finding 21:** There are roughly 3 groups in which the platforms can be divided. OSv (and OSv-FC, superimposed on top of OSv) and gVisor severely underperform. It is likely their custom thread allocators are to blame for this impaired performance, as these are the only two platforms that do not reuse existing mature thread implementations. The lack of any effect when varying the number of threads is also indicative of this. Moreover, as for gVisor, the high network latency undoubtedly worsens performance.

**Finding 22:** The second group consisting of gVisor and Firecracker yield performance around half of that of most other platforms. With complex real-world benchmarks like these, it remains difficult to say exactly precisely what causes this lower performance. For Firecracker, high memory latencies as demonstrated in the Memory benchmarks could be the root cause (as a subset of records is kept in memory during the benchmark). As for Kata containers, the relatively high I/O latency could be the culprit.

**Finding 23:** The third group, consisting of the remaining platforms, all perform alike. Due to the wide error bands (which did not narrow even when carrying out additional runs), there is no stable ranking of performance in this group.

## 4 Security and Isolation

Another critical property of any isolation platform is the degree of isolation it offers. We approximate this degree by measuring the horizontal attack profile (HAP), a term originally coined by IBM's James Bottomley [5]. The HAP is a quantitative metric that attempts to measure how wide the interface from the guest to the host is. Broadly speaking, the HAP is obtained by the amount of code executed multiplied by the bug density of the domain that is measured in. Concretely, this entails quantifying the number of host Linux kernel functions invoked while running different workloads in the guest. Multiplication of bug density is not needed as everything is measured within the same domain (i.e. the Linux kernel).

We extend the HAP metric by not only measuring how many functions are hit, but also taking into account *which* functions are hit. Functions get a score assigned based on their likelihood of exploitation, as obtained from the EPSS model [26]. This enables us to look at which functions are hit and then determine their exploitability, weighing functions that are more likely to be exploited heavier than those that have a lower likelihood of exploitability.

To determine this *extended HAP score* for each platform, we trace which (and the amount of) host kernel function invocations. The tracing is performed using `ftrace` with the `trace-cmd` front-end. The workloads run during tracing are the CPU, memory and I/O benchmarks from the Sysbench benchmarking suite [37], the iperf3 networking benchmark [24], and simply starting the platform and shutting it down after 1 minute. The results are shown in Figure 18.

There following important observations can be made:

**Finding 24:** Firecracker, the isolation platforms that advertises itself as running lightweight VMs and having a minimal device model, calls into the host kernel most often of all platforms. A wider interface between the host and guest is indicative of potentially weaker security. While the defense in depth is likely superior, Firecracker's minimalist approach still exposes a much wider interface to the host than the more heavy-weight general-purpose QEMU hypervisor.

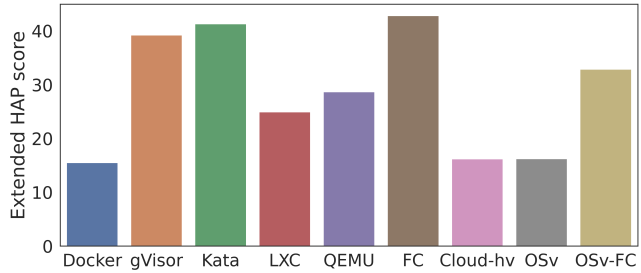

**Figure 18.** Results of the extended HAP metric

**Finding 25:** Cloud Hypervisor invokes very few function calls in the host kernel, which is surprising given the results of the other two hypervisors. As both the techniques as well as the architecture of Cloud Hypervisor are similar to Firecracker, this could be attributed to the work-in-progress status of the project, not fully supporting all functionality that the other hypervisors do.

**Finding 26:** The secure containers, gVisor and Kata container, have relatively high numbers, especially compared to the regular containers. For Kata containers this is to be expected, given that it starts its own hypervisor (albeit with a stripped down Linux kernel). The number of functions for gVisor is higher than expected; apparently the re-implementation of a kernel in user-space does not necessarily lead to fewer calls to the host kernel.

**Finding 27:** OSv, in particular given the fact that it uses a hypervisor, executes host kernel functions sparingly. OSv in this sense fulfills its promise of only exposing a narrow interface to the host, and more importantly, from this we can conclude that *a wide HAP is not inherent to the use of a hypervisor*.

**Finding 28:** The HAP metric fails to capture the defense-in-depth isolation platforms provide. For example, while Kata containers have a large HAP, they also introduce defense-in-depth, by using both namespaces as well as a hardware-assisted isolation mechanisms. Moreover, the potential attack surface between tenants of an isolation platform is also not accounted for. The HAP measures the (horizontal) width of the attack profile, but is unable to capture these vertical, defense-in-depth, aspects of the isolation platforms.

## 5 Discussion and Conclusions

Virtualization is the underlying technology that powers cloud infrastructure as we know it today. The increase in both mainstream adoption and variety of offerings underline the importance of several key properties of virtualization platforms. In particular, we consider the performance overhead and the degree of isolation offered of utmost importance. Prior research [7, 15] typically delved deep into either of these two properties, and are generally limited to a small set of examined platforms. This paper attempts to bridge this gap. It addresses both the performance and security aspects of various isolation platforms. We carried out an extensive collection of experiments, quantifying both the performance and security offered by the isolation platforms. These experiments include typical micro-benchmarks and real-world benchmarks, as well as a novel way to quantitatively measure the degree of isolation.

From our results, we can draw the following conclusions:

**Conclusion 1:** Container platforms, such as Docker and LXC, typically showcase near-native performance. Out of all the isolation platforms, containers perform the best, and typically also have a low start-up time.

**Conclusion 2:** Hypervisors always impose overhead in their networking and memory subsystems. Other subsystems, such as I/O and CPU do not necessarily exhibit this overhead, although it depends on the particular hypervisor that is used. Generally speaking, the more mature the hypervisor, the lower the overhead.

**Conclusion 3:** Secure containers show the weakest performance of all isolation platforms. The networking and memory subsystems perform near-native (as with hypervisors), but in particular I/O performance suffers. The primary reason for this is the use of network file systems, although improvements with the likes of virtio-fs are promising.

**Conclusion 4:** The OSv unikernel generally performs well, although its performance is hard to quantify due to instabilities and incompatibilities with the chosen benchmarks. Start-up times are comparable to containers.

**Conclusion 5:** Firecracker is not the fastest to boot in our experiments, unlike what is presented in [1].

**Conclusion 6:** The tagline of the Kata containers project "Speed of containers, security of VMs" generally does not hold in our experiments. Performance of various subsystems, with in particular I/O, is weak in comparison to hypervisors while the security under the HAP is not superior to conventional containers.

**Conclusion 7:** Pursuing the development of solutions and protocols that are specifically made for isolation platforms has proven to be fruitful. For example, using virtio-fs can bring performance gains for all isolation platforms.

**Conclusion 8:** Considering the degree of isolation offered by the various platforms, we found the OSv unikernel to exercise the least amount of code in the host kernel, closely followed by containers.

**Conclusion 9:** General purpose OSs running under hypervisors and particularly secure containers generally invoke the most host kernel functions. The latter is a particularly interesting observation, as the secure containers according to the HAP metric are deemed to be insecure; the catch being that these platforms primarily offer defense in depth.